\documentclass[manuscript]{aastex}

\newcommand{\Hipp}{{\it Hipparcos}}        

\newcommand{\HST}{{\it HST}}

\newcommand{\Teff}{T_{\rm eff}}

\newcommand{\kms}{{\>\rm km\>s^{-1}}}

\def\pmb#1{\setbox0=\hbox{#1}
  \kern-.02em\copy0\kern-\wd0
  \kern.01em\copy0\kern-\wd0
  \kern.01em\copy0\kern-\wd0
  \kern.01em\copy0\kern-\wd0
  \kern.01em\copy0\kern-\wd0
  \kern-.02em\raise.01em\box0 }

\def\ref#1#2{$^{#1}$}

\shorttitle{Age of HD 140283}
\shortauthors{Bond et al.}

\begin{document}

\title{HD~140283: A Star in the Solar Neighborhood that Formed Shortly After the
Big Bang\altaffilmark{1}}

\author{Howard E. Bond\altaffilmark{2,3,4},
Edmund P. Nelan\altaffilmark{2},
Don A. VandenBerg\altaffilmark{5}, 
Gail H. Schaefer\altaffilmark{6},
and Dianne Harmer\altaffilmark{7}
}

\altaffiltext{1}
{Based in part on observations made with the NASA/ESA {\it Hubble Space
Telescope}, obtained by the Space Telescope Science Institute. STScI is operated
by the Association of Universities for Research in Astronomy, Inc., under NASA
contract NAS5-26555.}

\altaffiltext{2}
{Space Telescope Science Institute, 
3700 San Martin Dr.,
Baltimore, MD 21218, USA;
bond@stsci.edu, nelan@stsci.edu}

\altaffiltext{3}
{Department of Astronomy \& Astrophysics, Pennsylvania State University,
University Park, PA 16802, USA}

\altaffiltext{4} 
{Current address: 9615 Labrador Lane, Cockeysville, MD 21030, USA}

\altaffiltext{5}
{Dept.\ of Physics \& Astronomy, University of Victoria, P.O. Box 3055,
Victoria, BC, V8W 3P6, Canada; vandenbe@uvic.ca}

\altaffiltext{6}
{The CHARA Array of Georgia State University, Mount Wilson Observatory, Mount
Wilson, CA 91023, USA; schaefer@chara-array.org}

\altaffiltext{7}  
{National Optical Astronomy Observatories, 950 North Cherry Avenue, Tucson, AZ
85726, USA; diharmer@noao.edu}

\begin{abstract}

HD\,140283 is an extremely metal-deficient and high-velocity subgiant in the
solar neighborhood, having a location in the HR diagram where absolute magnitude
is most sensitive to stellar age. Because it is bright, nearby, unreddened, and
has a well-determined chemical composition, this star avoids most of the issues
involved in age determinations for globular clusters. Using the Fine Guidance
Sensors on the {\it Hubble Space Telescope}, we have measured a trigonometric
parallax of $17.15\pm0.14$~mas for HD\,140283, with an error one-fifth of that
determined by the {\it Hipparcos\/} mission. Employing modern theoretical
isochrones, which include effects of helium diffusion, revised nuclear reaction
rates, and enhanced oxygen abundance, we use the precise distance to infer an
age of $14.46\pm0.31$~Gyr. The quoted error includes only the uncertainty in the
parallax, and is for adopted surface oxygen and iron abundances of
$\rm[O/H]=-1.67$ and $\rm[Fe/H]=-2.40$.   Uncertainties in the stellar
parameters and chemical composition, especially the oxygen content, now
contribute more to the error budget for the age of HD\,140283 than does its
distance, increasing the total uncertainty to about $\pm$0.8~Gyr. Within the
errors, the age of HD\,140283 does not conflict with the age of the Universe,
$13.77\pm0.06$~Gyr, based on the microwave background and Hubble constant, but
it must have formed soon after the big bang.

\end{abstract}





\section{Introduction: The Oldest Stars}

The age of the Universe is $13.77\pm0.06$~Gyr, based on data on the cosmic
microwave background (CMB), baryon acoustic oscillations, and Hubble constant
(Bennett et al.\ 2012). Precise ages for the oldest and most metal-deficient
stars can date the onset of star formation (e.g., Bromm \& Larson 2004)
following the big bang. Moreover, because the oldest stars must be younger than
the Universe, precise ages provide a strong test of the consistency of stellar
and cosmological physics. 

As recently as the 1990's, there appeared to be a conflict between relatively
high ages found for stars in Galactic globular clusters (GCs), and a relatively
low age of the Universe from determinations of the Hubble constant. This
situation changed with the discovery of evidence for an accelerating expansion
of the Universe (Riess et al.\ 1998; Perlmutter et al.\ 1999), improved
precision in determinations of the Hubble constant (Freedman et al.\ 2001), and
measurements of the CMB\null. At about the same time, the first studies to use
parallaxes of local subdwarf calibrators from the \Hipp\/ mission (Perryman et
al.\ 1997) to derive GC distances suggested that these systems are younger than
previously thought (Reid 1997; Gratton et al.\ 1997; Chaboyer et al.\ 1998; but
see Pont et al.\ 1998).  However, the increased cluster distance moduli reported
in these papers, which were responsible for the reduced ages, tended to be
$\sim$0.2~mag larger than currently favored estimates (as given in the latest
version of the online catalog maintained by Harris 1996).

By the early 2000s, refinements in stellar-evolution modeling (most notably the
inclusion of diffusive processes) again reduced GC ages to values that appeared
to be less than the age of the Universe (VandenBerg et al.\ 2002), only to be
followed by a revision of the rate of the $^{14}{\rm N}(p,\,\gamma)^{15}\rm O$
reaction, which implied increased ages at a given turnoff luminosity by
0.7--1.0~Gyr (Imbriani et al.\ 2004). Although it is unlikely that there will be
further significant revisions to basic stellar physics, because most of the
ingredients of stellar models have been carefully examined during the past
decade, it is still unclear whether or not GC ages are compatible with the age
of the Universe. In addition to the issue of distances, age determinations of
stars in GCs require knowledge of interstellar extinction and chemical
compositions, including the abundances of individual heavy elements.

An alternative approach to stellar chronology is to determine ages of extreme
Population~II subgiants in the solar neighborhood based on direct trigonometric
parallaxes, combined with state-of-the-art theoretical isochrones appropriate to
the detailed composition of each star. This method bypasses most of the problems
associated with the much more distant clusters. 

\section{The Extreme Halo Subgiant HD 140283}

The ideal solar-neighborhood target for an age determination based on a precise
parallax would be a nearby extremely metal-deficient star, with a
well-determined chemical composition based on high-resolution spectroscopy,
which has begun to evolve off the main sequence in the Hertzsprung-Russell
diagram (HRD)\null. The one star that best satisfies these criteria is
HD~140283, a bright (7th-mag) Population~II subgiant with a very low metal
content ($\rm[Fe/H]=-2.40\pm0.10$; Casagrande et al.\ 2010). HD~140283 played an
important role in astronomical history as the first high-velocity star
recognized---a century ago---to have an anomalously early spectral type for its
low luminosity (Adams 1912), making it a so-called ``A-type subdwarf.'' 
Moreover, HD~140283  (along with another similar subdwarf, HD\,19445) was
subsequently the first star shown, through spectroscopic analysis, to have a
much lower heavy-element content than the Sun (Chamberlain \& Aller 1951; see
also the historical discussion by Sandage 2000). This explains the superficial
resemblance of its extremely weak-lined spectrum to that of a hotter star of
normal composition, in spite of a surface temperature corresponding to an early
G-type star. Thus HD~140283 was the key to the realization that the chemical
elements heavier than helium are synthesized during stellar evolution (Burbidge
et al.\ 1957), making low abundances of the heavy elements a hallmark of the
oldest stars. With improved photometry, trigonometric parallaxes, and
spectroscopic analyses, it was recognized later (Cohen \& Strom 1968; Cayrel
1968) that HD~140283 is actually a slightly evolved subgiant, rather than a
classical subdwarf, placing it at the ideal location in the HRD where the
absolute magnitude is most sensitive to stellar age.

The \Hipp\/ parallax of HD~140283 is $17.16\pm0.68$~mas, according to a recent
re-analysis of the \Hipp\/ data (van Leeuwen 2007).  The corresponding
luminosity, combined with isochrones calculated without element diffusion,
implied an age for the star greater than $\sim$14~Gyr; and if HD~140283 were
used to calibrate the distances of GCs, the oldest clusters were found to have
ages of at least 15~Gyr (VandenBerg 2000).  Inclusion of effects of helium
diffusion in the calculations reduced the implied age of HD\,140283 to
$13.5\pm1.5$~Gyr (VandenBerg et al.\ 2002), but still with an uncomfortably
large uncertainty. The largest contributor to this relatively large error in the
age is the uncertainty in the \Hipp\/ parallax.  As noted in \S1, the revision
of the $^{14}$N$(p,\,\gamma)^{15}$O reaction that occurred two years later would
have increased the predicted age to $\sim$14.3~Gyr, if all of the other factors
that play a role in the age determination were left unchanged.

\section{Hubble Space Telescope FGS Astrometry of HD~140283}

At the present time, the most precise trigonometric parallaxes that can be
obtained at optical wavelengths are from the Fine Guidance Sensors (FGS) on the
{\it Hubble Space Telescope\/} (\HST\/)\null. The FGS have been shown to be
capable of yielding parallaxes with better than 0.2~mas precision (e.g.,
Benedict et al.\ 2007). Because of the importance of the ages of the oldest
stars, we undertook observations of HD~140283 aimed at improving the precision
of its parallax, and reducing any systematic errors, relative to the \Hipp\/
result. 

We made FGS observations of HD\,140283 at 11 epochs between 2003 August and 
2011 March, at dates close to the biannual times of maximum parallax factor. 
The FGS are interferometers that, in addition to providing guiding control
during imaging or spectroscopic observations, can measure precise positions of a
target star and several surrounding astrometric reference stars with one FGS
while the other two guide the telescope.  These positional measurements are
corrected for differential velocity aberration, geometric distortion, thermally
induced spacecraft drift, and jitter.  Because of refractive elements in the FGS
optical train, an additional correction based on the $B-V$ color of each star is
applied. Moreover, due to its brightness, HD\,140283 itself was observed with
the F5ND neutral-density attenuator, while the reference stars were observed
only with the F583W filter element.  Thus it was necessary to apply a
``cross-filter" correction to the positions of HD\,140283 relative to the
reference stars.

The data from all epochs were combined using a four-parameter (translation in
$x$ and $y$, rotation, and scale) overlapping-plate technique to form a master
plate. We employed the least-squares program GAUSSFIT (Jefferys, Fitzpatrick, \&
McArthur 1988) to solve for the parallax and proper motion of the target and the
six reference stars, as outlined in detail by Benedict et al.\ (2011). Since the
FGS measurements provide only the relative positions of the stars, the model
requires input estimated values of the reference-star parallaxes and proper
motions, in order to determine an absolute parallax of the target. These
estimates (see next paragraph) were input to the model as observations with
errors, which permits the model to adjust their parallaxes and proper motions
(to within their specified errors) to find a global solution that minimizes the
resulting $\chi^2$ fit.

As just noted, the solutions for parallax and proper motion of the target star
require estimates of the distances and proper motions of the background
reference stars. We made the distance estimates using ground-based spectroscopy
and photometry of the six reference stars (whose $V$ magnitudes range from 11.9
to 16.6). Due to space limitations, the details of this process will be
published elsewhere, but we summarize here. For spectral classification, we
obtained digital spectra with the WIYN 3.5m telescope and Hydra spectrograph at
Kitt Peak National Observatory, and with the 1.5m SMARTS telescope and
Ritchey-Chretien spectrograph at Cerro Tololo Interamerican Observatory
(CTIO)\null. The classifications were then accomplished through comparison with
a network of standards obtained with the same telescopes, assisted by
equivalent-width measurements of lines sensitive to temperature and luminosity.

Photometry of the reference stars in the Johnson-Kron-Cousins {\it BVI\/} system
was obtained with the SMARTS 1.3m telescope at CTIO, using the ANDICAM CCD
camera, and calibrated to the standard-star network of Landolt (1992). Each star
was observed on five different photometric nights in 2003, 2005, and 2007.  To
estimate the reddening of the reference stars (assumed to be the same for all
six, since their distances place them well beyond the dust of the Galactic disk
and also well behind the unreddened HD\,140283 itself), we compared the observed
$B-V$ color of each star with the intrinsic $(B-V)_0$ color corresponding to its
spectral type (Schmidt-Kaler 1982), and calculated the average $E(B-V)$\null. We
also used the extinction map of Schlafly \& Finkbeiner (2011), as implemented at
the NASA/IPAC website\footnote{\url{http://ned.ipac.caltech.edu}}, to determine
the reddening in the direction beyond HD~140283. Both methods yielded
$E(B-V)=0.14$, which was used to correct all of the magnitudes and colors. 
Finally, we estimated the distances as follows: (1)~For stars classified as
subgiants and giants, we fitted them by interpolation to a fiducial sequence
[$M_V$ vs.\ $(V-I)_0$] for the old open cluster M67 (Sandquist 2004). (2)~For
the stars classified as dwarfs, we derived calibrations of the visual absolute
magnitude, $M_V$, against $B-V$ and $V-I$ colors through polynomial fits to a
sample of 791 single main-sequence stars with accurate {\it BVI\/} photometry
and \Hipp\/ or USNO parallaxes of 40~mas or higher ($d<25$~pc), which is
provided online by
I.~N.~Reid\footnote{\url{http://www.stsci.edu/$\sim$inr/cmd.html}}. A correction
for metallicity, estimated from each star's position in $B-V$ vs.\ $V-I$, was
applied.  We tested our algorithm by applying it to 136 nearby stars with
accurate parallaxes and a wide range of metallicities listed by Casagrande et
al.\ (2010). We reproduced their known absolute magnitudes with an rms scatter
of only 0.28~mag. At the distances of the reference stars, ranging from 650 to
1700~pc, this scatter corresponds to parallax errors of 0.08 to 0.20~mas.

The initial proper-motion estimates for the reference stars were taken from two
independent catalogs: UCAC4 (Zacharias et al.\ 2012) and PPMXL (Roeser,
Demleitner, \& Schilbach 2010). We noted that the proper motions disagreed
between the UCAC4 and PPMXL catalogs for several of our reference stars by more
than the stated errors. Thus we ran two independent solutions based upon the two
catalogs. To minimize contamination of the FGS results by large input catalog
errors, we applied an iterative technique whereby the FGS proper motions that
were output from the solution using either UCAC4 and PPMXL as input were used as
the input proper motions in a second iteration. This resulted in good agreement
($\sim$$1\,\rm mas\,yr^{-1}$) between the relative proper motions of the
reference stars, but a systematic difference in the absolute values. The
solution based upon the UCAC4 catalog reproduced the \Hipp\/ proper motion
of HD\,140283 quite well, with the solution based on PPMXL differing by
$\sim$$3\,\rm mas\,yr^{-1}$. Both models yielded parallaxes of HD\,140283 that
agreed within 0.03~mas.

For our final solution, we used the UCAC4 catalog and the iterative procedure
described above. The resulting absolute parallax of HD~140283 is
$17.15\pm0.14$~mas ($d=58.30\pm0.48$~pc). The uncertainty includes contributions
from residual errors in the geometric-distortion calibration of the FGS, errors
in \HST\/ pointing performance, and errors in the raw stellar position
measurements.  The resulting proper-motion components for HD\,140283 from the
FGS solution are $(\mu_\alpha,\mu_\delta) = (-1114.50\pm0.12,
-304.59\pm0.11)\,\rm mas\,yr^{-1}$, which agree very well with the absolute
proper motion determined by \Hipp, $(-1114.93\pm0.62, -304.36\pm0.74)\,\rm
mas\,yr^{-1}$. The tangential velocity of HD\,140283 is thus $319.3\kms$, and
its total space motion relative to the Sun, taking into account the radial
velocity of $-169.0\kms$, is $361.3\kms$.

\section{The Age of HD 140283}

Our FGS parallax for HD\,140283, together with an apparent visual magnitude $V =
7.205\pm0.02$ (Casagrande et al.\ 2010) and $E(B-V)=0.000\pm0.002$ (Mel\'endez
et al.\ 2010), yields a visual absolute magnitude $M_V = +3.377 \pm 0.027$. A
recent calibration of the infrared-flux method gives an effective temperature
$\Teff=5777 \pm 55$~K (Casagrande et al.\ 2010). If this $\Teff$ is adopted in
high-resolution spectroscopic analyses (Gratton, Carretta, \& Castelli 1996;
Israelian et al.\ 2004; Rich \& Boesgaard 2009), the resulting iron abundance
relative to hydrogen is $\rm[Fe/H] = -2.40 \pm 0.10$, where values in square
brackets are logarithms of the abundances by number, normalized to the solar
values.  According to stellar models that allow for diffusive processes and
extra (turbulent) mixing to limit the efficiency of gravitational settling
(Richard et al.\ 2002), the [Fe/H] measured at the stellar surface will be
$\sim$0.1~dex lower than its value in the interior. We therefore adopt
$\rm[Fe/H]=-2.3$ for our interior stellar models of HD\,140283.

To determine the age of HD$\,$140283, we employed evolutionary tracks and
isochrones computed using the current version of the University of Victoria code
(VandenBerg et al.\ 2012), with an adopted helium abundance by mass of $Y =
0.250$, slightly above recent estimates of the primordial He abundance,
$Y_0=0.2486$ (Cyburt, Fields, \& Olive 2008). The Victoria models take into
account current values for nuclear-reaction rates (see \S\S1--2), and include
the diffusive settling of helium. Diffusion of elements heavier than He is not
treated, apart from the small adjustment to $\rm[Fe/H] = -2.3$ described above,
but the effect of this neglect on derived ages is very small. 

At low metallicities, the locations of the turnoff and subgiant portions of
isochrones in the HRD depend most strongly on the absolute abundance of oxygen
(e.g., VandenBerg et al.\ 2012), and less so on the abundance of iron and other
heavy metals. For example, if [O/H] is fixed at $-1.4$, nearly the same
isochrones are obtained if $\rm[Fe/H] = -2.3$ and $\rm[O/Fe] = +0.9$, or if
$\rm[Fe/H] = -1.9$ and $\rm[O/Fe] = +0.5$. Unlike oxygen, helium is predicted to
have almost no impact on the location of the subgiant branch in the HRD, as
first shown by Carney (1981; see also VandenBerg et al.\ 2012, Fig.~18).  Thus
the age of HD\,140283 is essentially independent of $Y$. This is fortunate
because helium lines are not detectable in the spectra of cool stars, and thus
the abundance of He cannot be measured directly.

The crucial oxygen abundances in metal-poor G-type stars like HD\,140283 can be
determined using three different spectroscopic features: the forbidden
[\ion{O}{1}] line at 6300~\AA, the high-excitation permitted \ion{O}{1} triplet
at 7771--7775~\AA, or molecular OH bands in the optical ultraviolet or infrared.
As reviewed by many authors (e.g., Asplund 2005; Mel\'endez et al.\ 2006;
Fabbian et al.\ 2009b), each of these methods have difficulties: the
[\ion{O}{1}] line is extremely weak and potentially blended; the permitted
\ion{O}{1} lines are affected by departures from local thermodynamic
equilibrium; and the OH bands are subject to problems of three-dimensional
effects in the stellar atmosphere (i.e., granulation), blends, and continuum
placement.

Recent determinations of the [O/H] ratio in HD\,140283 using all three of these
methods (for example, Nissen et al.\ 2002; Mel\'endez et al.\ 2006; Rich \&
Boesgaard 2009; Tan et al.\ 2009) have ranged from [O/H] = $-1.55$ to $-1.78$
(corrected to the temperature adopted here of $\Teff=5777$~K using the parameter
dependencies given by the cited authors), with errors typically of order
$\pm$0.1--0.15~dex. The unweighted mean of these authors' determinations is
about $\rm[O/H]=-1.67$, corresponding to $\rm[O/Fe]\simeq+0.7$. Corrected for
diffusion, the initial oxygen abundance would have been higher by $\sim$0.13~dex
(Richard et al.\ 2002, their Fig.~13), resulting in $\rm[O/H]_0=-1.54$.  We
convert this to the absolute O abundance by using a solar value of $\log N({\rm
O})=8.69\pm0.05$ (Asplund et al.\ 2009).

The top panel in Figure~1 shows the position of HD\,140283 in the
semi-theoretical HRD ($M_V$ vs.\ $\log\Teff$) along with isochrones derived for
ages of 13.4, 13.9, and 14.4~Gyr.  The implied age is 14.46~Gyr. The uncertainty
in the age due only to the errors in parallax, photometry, and extinction is
only $\pm$0.31~Gyr,  about one-fifth of that associated with previous age
determinations based on the \Hipp\/ parallax (VandenBerg et al.\ 2002). Other
contributors to the error budget are summarized in Table~1, and are due to
uncertainties in the stellar parameters and composition, dominated by the
uncertainty in the [O/H] ratio. The total error in the age is about $\pm$0.8~Gyr
(calculated by combining the errors independently in quadrature; this is an
approximation because there may be correlations between, e.g., errors in $\Teff$
and derived abundances; we did not attempt to model these, since [O/H] is the
dominant source of uncertainty). 

The age of HD\,140283 does not conflict with the age of the Universe, given the
$\pm$0.8~Gyr uncertainty. The middle and lower panels in Figure~1 illustrate the
sensitivity of the age to an increased oxygen abundance. If [O/H] is increased
by 0.15~dex, which is roughly the uncertainty in the measured abundance, the age
of HD\,140283 is reduced to about 13.8~Gyr. Increasing [O/H] by 0.30~dex reduces
the age to $\sim$13.3~Gyr. 

In summary, the age determination for HD\,140283 depends primarily on its oxygen
abundance relative to hydrogen, with its parallax now having been removed as an
important contributor to the error budget. The most important reasons for
differences between our new result and that of VandenBerg et al.\ (2002) are the
inclusion of diffusion, new nuclear reaction rates, adoption of $E(B-V) = 0.00$
instead of 0.025, and the much more precise parallax.

\section{Implications}

HD$\,$140283 is the oldest known star for which a reliable age has been
determined---but of course it is not quite a primordial star, given its low but
non-zero metallicity. Our precise distance could enable HD$\,$140283 to be used
as a ``standard candle" to determine the distances to very metal-deficient GCs
and nearby metal-poor dwarf galaxies (which appear to be coeval with GCs---Brown
et al.\ 2012), by fitting their subgiant branches to the luminosity of
HD\,140283. However, this would result in a distance to the GC M92 smaller by
$\sim$0.1--0.2~mag than nearly all other distance estimates. This discrepancy
would be reduced (and the implied age of HD\,140283 also reduced) if the star
is slightly reddened. Setting $E(B-V)=0.02$, in spite of the zero value found by
Mel\'endez et al.\ (2010), would reduce its age by $\sim$0.65~Gyr. 


There is a remarkable accordance (within their respective uncertainties) between
the age of the Universe inferred from the CMB, the age of the chemical elements
(Roederer et al.\ 2009), and the ages of the oldest stars. The difficulty of
determining accurate abundances, especially [O/H], will continue to limit the
accuracy of stellar age determinations for the foreseeable future, including the
era when accurate distances out to a few kpc are obtained from {\it Gaia\/}
(Perryman et al.\ 2001).

\acknowledgments

We thank STScI for support through grant GO-9883.  H.E.B. was Visiting
Astronomer, Kitt Peak National Observatory, National Optical Astronomy
Observatory, which is operated by the Association of Universities for Research
in Astronomy, Inc., under cooperative agreement with the National Science
Foundation.  D.A.V. acknowledges a Discovery Grant from the Natural Sciences and
Engineering Research Council of Canada. Based in part on observations obtained
with the SMARTS Consortium 1.3- and 1.5-m telescopes located at CTIO, Chile.
STScI summer students Ryan Leaman and Mihkel Kama assisted in data reduction for
the reference stars. We thank B. E. McArthur for calibration of the FGS1
geometric distortion. 

{\it Facilities:} \facility{Hubble Space Telescope, SMARTS 1.3- and 1.5-m
telescopes, WIYN 3.5-m telescope}


\begin{figure}
\begin{center}
\includegraphics[width=6.5in]{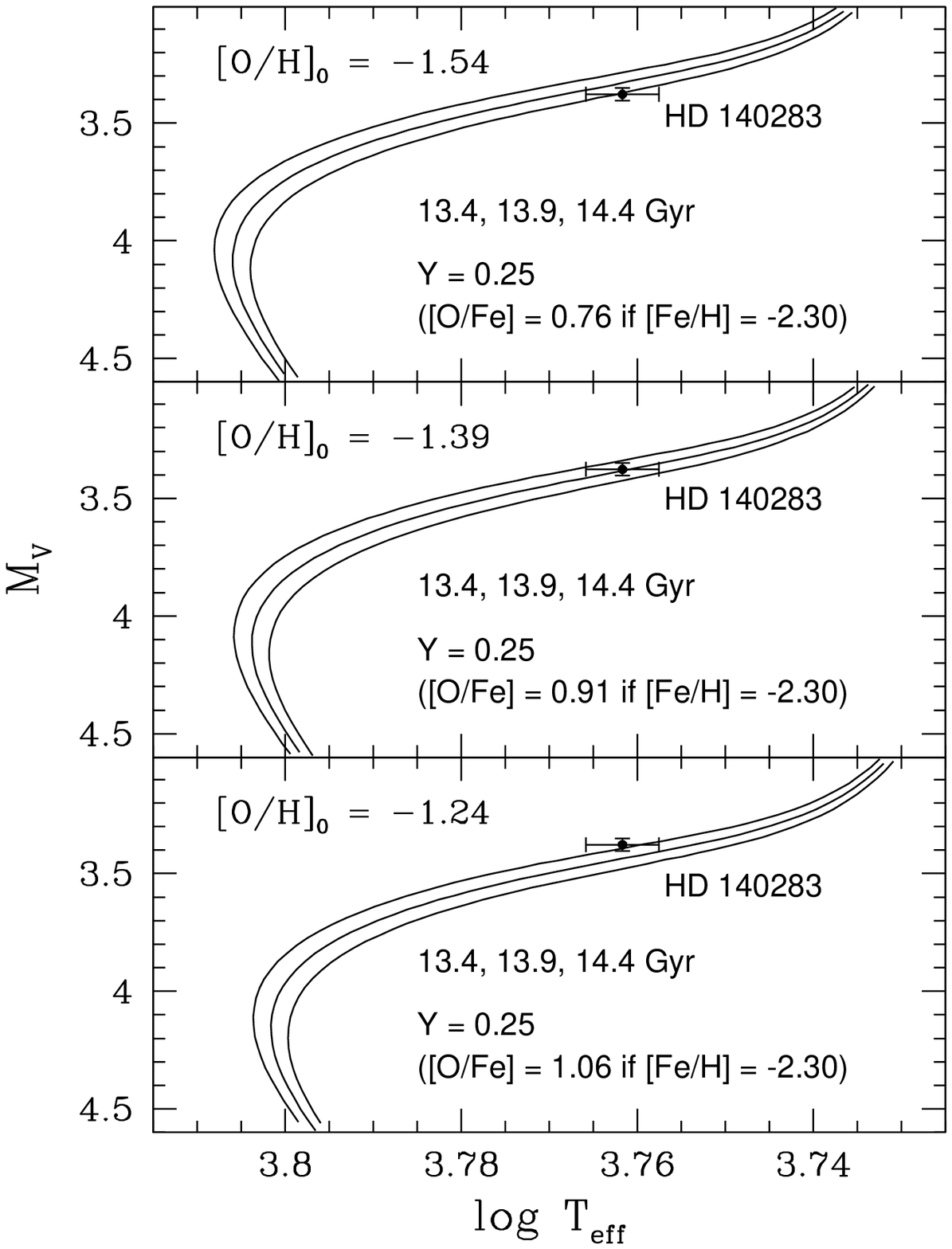}
\end{center}
\end{figure}

\begin{figure}
\begin{center}
\figcaption{
{\it Top panel:} Theoretical University of Victoria isochrones, plotted in the
$(\log T_{\rm eff},\,M_V)$ plane, for stellar ages of 13.4, 13.9, and 14.4~Gyr.
The plot is magnified to show only the subgiant portion, lying between the
main-sequence turnoff to the lower left and the base of the giant branch to the
upper right. For the composition in the stellar interior, we used a helium
abundance of $Y=0.25$, an iron abundance of $\rm[Fe/H]=-2.3$, and an initial
oxygen abundance of $\rm[O/H]=-1.54$. The point with error bars shows the
location of HD$\,$140283. The implied age is $14.46\pm0.31$~Gyr, where the error
bar is the uncertainty due only to the error in the {\it Hubble Space
Telescope\/} trigonometric parallax.  The systematic error in the age due to
uncertainties in the effective temperature and chemical composition is larger,
about $\pm$0.8~Gyr (see text).
{\it Middle and lower panels:} Theoretical isochrones for ages of 13.4, 13.9,
and 14.4~Gyr, with the initial [O/H] abundance ratio increased by 0.15 and
0.30~dex, decreasing the implied age of HD\,140283 to about 13.8 and 13.3~Gyr,
respectively.  The uncertainty in the initial oxygen abundance is now the
largest contributor to the uncertainty in the age.}
\end{center}
\end{figure}

\clearpage

\begin{deluxetable}{lllc}
\tablewidth{0 pt}
\tablecaption{Error Budget for Age of HD\,140283}
\tablehead{
\colhead{Quantity} &
\colhead{Value} &
\colhead{Uncertainty} &
\colhead{$\sigma(\rm age)$ [Gyr] }}
\startdata
Parallax                & 17.15   & $\pm$0.14 mas & 0.21 \\
$V$                     & 7.205   & $\pm$0.02 mag & 0.23 \\
$E(B-V)$                & 0.000   & $\pm$0.002 mag& 0.06 \\
$\Teff$                 & 5777    & $\pm$55 K     & 0.35 \\
\leavevmode[Fe/H]       & $-2.40$ & $\pm$0.10 dex & 0.10 \\
\leavevmode[O/H]        & $-1.67$ & $\pm$0.15 dex & 0.61 \\
Solar $\log N({\rm O})$ & 8.69    & $\pm$0.05 dex & 0.20 \\
\noalign{\vskip0.1in}
Total uncertainty       &         &               & 0.80 \\
\enddata
\tablecomments{Sources of values and error estimates are given in text.}
\end{deluxetable}

\end{document}